\documentclass[aps,prl,preprint,superscriptaddress]{revtex4-2}

\usepackage{graphicx}
\usepackage{bm}
\usepackage{longtable}

\usepackage{graphicx}
\usepackage{dcolumn}
\usepackage{bm}

\begin{document}

\preprint{00000}

\title{Observing the effects of numbers of valence nucleons on $0_{gs}^+ \rightarrow 2_1^+$ transitions in deformed nuclei by comparing proton and neutron transition matrix elements} 

\author{P. D. Cottle} \affiliation{Department of Physics, Florida State University, Tallahassee, FL 32306, USA}

\author{L.A. Riley}
\affiliation{Department of Physics and Astronomy, Ursinus College, Collegeville, PA 19426}%

\author{A. Gade} \affiliation{Facility for Rare Isotope Beams, Michigan State University, East Lansing, MI, 48824, USA}
\affiliation{Department of Physics and Astronomy, Michigan State University, East Lansing, MI, 48824, USA}

\author{K. W. Kemper} \affiliation{Department of Physics, Florida State University, Tallahassee, FL 32306, USA}

\author{M. Spieker} \affiliation{Department of Physics, Florida State University, Tallahassee, FL 32306, USA}

\date{\today}

\begin{abstract}

We examined the ratios of neutron and proton transition matrix elements, $M_n/M_p$, for the $0_{gs}^+ \rightarrow 2_1^+$ transitions in 48 even-even stable nuclei with $N>20$ for which electromagnetic matrix elements were compiled by Pritychenko \textit{et al.} and for which high-quality inelastic proton scattering data were available.  Several deformed rare earth nuclei have $(M_n/M_p)/(N/Z)$ values significantly below 1.0, which is not consistent with a simple liquid drop picture.  However, this phenomenon can be explained using a schematic picture in which $M_p$ reaches a maximum at proton mid-shell ($Z=66$) and $M_n$ reaches its maximum at neutron midshell ($N=104$).  Several mid-mass vibrational nuclei have $M_n/M_p$ values significantly below $N/Z$, which is not consistent with the expectation that $M_n/M_p = N/Z$ in such nuclei.  A shell model investigation of these observations might yield insights about this behavior.

\end{abstract}

\maketitle


A simple liquid drop model of deformed nuclei intrinsically assumes that the neutron and proton fluids are homogeneously distributed throughout the nucleus.  In the simplest liquid drop picture, the radii of the proton and neutron fluids are equal and have equal deformation parameters.  A somewhat more refined picture would allow the protons and neutron fluids to have different radii, depending on the proton and neutron numbers. 

A recent study of the $0_{gs}^+ \rightarrow 2_1^+$ transition in the deformed neutron-rich nucleus $^{42}$Si with both intermediate-energy Coulomb excitation and inelastic proton scattering in inverse kinematics \cite{Ri25} to determine the ratio of the proton and neutron transition matrix elements, $M_n/M_p$, found that neither of these simple liquid drop pictures can account for the experimental results, but that a shell model calculation can.  


In the present work, we followed up on the $^{42}$Si result by examining the ratios of neutron and proton transition matrix elements, $M_n/M_p$, for $0_{gs}^+ \rightarrow 2_1^+$ transitions in 48 even-even stable isotopes for which both electromagnetic and high quality inelastic proton scattering data are available.  We examined two sets of such nuclei.  One set is composed of nuclei in the neutron number range $N=82-126$, which includes the deformed rare earth region.  Our survey of this region allows an examination of how $M_n/M_p$ values evolve as nuclear shapes transition from spherical to vibrational to deformed.  The second set of nuclei we examined are those in other mass regions that, according to the compilation of $B(E2;0_{gs}^+ \rightarrow 2_1^+)$ values in Ref. \cite{Pri16}, have electromagnetic quadrupole deformation parameters $\beta_2$ of 0.20 or greater.  This set includes both transitional and deformed isotopes.  However, we only included nuclei with $N>20$ to avoid clustering effects.  The intent of our analysis was to determine to what extent the $M_n/M_p$ values for $0_{gs}^+ \rightarrow 2_1^+$ transitions in even-even nuclei with stable deformation can be accounted for within the two liquid drop pictures described in the first paragraph.  

The use of two complementary experimental probes for measuring $0_{gs}^+ \rightarrow 2_1^+$ transitions allows the determination of the ratio of the neutron and proton transition matrix elements, $M_n/M_p$, where $M_n$ and $M_p$ are given by \cite{Be81,Be83}

\begin{equation}
\label{eq:definition2}
M_{n(p)} = \int_0^{\infty} \rho_{fi}^{n(p)}(r)r^{\lambda + 2} \  dr 
\end{equation}

\noindent and where $\rho_{fi}^{n(p)} (r)$ is the neutron (proton) transition density and $\lambda$ is the multipolarity of the transition.  An electromagnetic measurement is only sensitive to the proton contribution to the transition, so it provides a measurement of $M_p$.  Inelastic proton scattering is sensitive to contributions from both protons and neutrons, so the combination of the two probes allows the calculation of  $M_n/M_p$ \cite{Be83}.

The ratio $M_n/M_p$ provides information about both the nucleus itself and the nature of the excitation measured.  For example, Bernstein, Brown and Madsen pointed out \cite{Be81,Be83} that for the excitation of an isoscalar nuclear vibration - that is, one in which the proton and neutron fluids vibrate with equal amplitudes and in phase - the neutron and proton transition matrix elements are related by $M_n/M_p = N/Z$.  They demonstrated that the neutron and proton transition matrix elements for the $0_{gs}^+ \rightarrow 2_1^+$ excitations in many even-even open shell vibrational nuclei conform to this relationship \cite{Be81,Be83}.

The relationship $M_n/M_p = N/Z$ that holds for a vibrational nucleus in a liquid drop model also applies to a stably deformed nucleus, although prior to the present work no systematic comparison of $M_n/M_p$ values in deformed nuclei to the liquid drop expectation has been performed.   

Bernstein, Brown and Madsen also focused on stable even-even single closed shell (SCS) nuclei.  In such isotopes, the $2_1^+$ states are often dominated by promotions of valence nucleons from one orbit to another.  If core polarization played no role in the $0_{gs}^+ \rightarrow 2_1^+$ excitations of these nuclei, then such excitations in closed-neutron-shell nuclei would have $M_n=0$ and in closed proton shell nuclei $M_p=0$.  In actual SCS nuclei, core polarization occurs and $M_n$ and $M_p$ always have non-zero values.  Nevertheless, for $0_{gs}^+ \rightarrow 2_1^+$ excitations in closed-proton-shell even-even nuclei that have valence neutrons the ratio $M_n/M_p$ is systematically larger than $N/Z$.  Likewise, in closed-neutron-shell even-even nuclei that have valence protons, $M_n/M_p$ is systematically smaller than $N/Z$ \cite{Be81,Be83,Ke92}.


Bernstein, Brown and Madsen devised a prescription (which we will call the ``Bernstein prescription") for extracting the $M_n/M_p$ value from the $B(E2;0_{gs}^+ \rightarrow 2_1^+)$ electromagnetic value and the deformation length extracted from the inelastic proton scattering reaction, $\delta_{(p,p')}$, that assumes the simpler of the two liquid drop pictures described in the opening paragraph of this work. That simpler liquid drop picture is that the neutrons and protons are homogeneously distributed in the nucleus and that the radii of the neutron and proton fluids are identical.

Bernstein, Brown and Madsen \cite{Be83} described their prescription in the following way.  First, a proton deformation length, $\delta_p$, is determined from the $B(E2;0_{gs} \rightarrow 2_1^+)$ value using 

\begin{equation}
\label{eq:delta_p}
\delta_p = \frac{4 \pi}{3 Z} \frac{1}{r_C A^{1/3}} \sqrt{\frac{B(E2; 0^+_{g.s.} \rightarrow 2^+_1)}{e^2}} 
\end{equation}

\noindent
where $r_C = 1.20$~fm is the radius parameter of the Coulomb potential.

With these values of $\delta_p$ and $\delta_{(p,p')}$, a rearrangement of Equation (7) in \cite{Be83} gives,

\begin{equation}
\label{eq:mnmp2}
\frac{M_n}{M_p} = \frac{b_p}{b_n} \left[\frac{\delta_{(p,p')}}{\delta_p} \left( 1 + \frac{N}{Z} \frac{b_n}{b_p} \right) - 1 \right],
\end{equation}

\noindent
where $\frac{b_n}{b_p}$ is the ratio of the sensitivities of the proton inelastic scattering probe to neutrons and protons.  

According to Bernstein, Brown and Madsen \cite{Be81}, the value of $\frac{b_n}{b_p}$ for inelastic proton scattering varies with the incident energy of the proton.  They said that at low incident energies, $\frac{b_n}{b_p}=3$.  De Leo \textit{et al.} \cite{deL96} argued that the $\frac{b_n}{b_p}=3$ value could be used for incident energies as high as 65 MeV.  De Leo \textit{et al.} also said that $\frac{b_n}{b_p}$ is ``assumed to be proportional to the volume integrals of the central isoscalar and isovector potentials".  Finley \textit{et al.} \cite{Fi80} concluded that there is a $25\%$ uncertainty in isovector parameters in nucleon inelastic scattering.  This implies that there is a $25\%$ uncertainty in $\frac{b_n}{b_p}$ for low energy (up to 65 MeV) proton scattering as well, giving $\frac{b_n}{b_p}=3.00 \pm 0.75$.  

Bernstein, Brown and Madsen also reported \cite{Be81} that at a proton incident energy of 800 MeV, $\frac{b_n}{b_p}=0.83$.  At incident energies between these limits, there is less certainty about the value of $\frac{b_n}{b_p}$.


In contrast to the Bernstein prescription, the prescription for calculating $M_n/M_p$ from the results of complementary probes developed by Khan \cite{Kh22} (which we will call the ``Khan prescription") assumes the more refined of the two liquid drop pictures described in the opening paragraph of this work.  That is, the Khan prescription allows the radii of the neutron and proton fluids to be different.  In this way, it is intended to account for proton and neutron excesses in exotic nuclei more accurately than the Bernstein prescription does.  In addition, it takes into account the diffuseness of the nucleon fluids at their surfaces and allows the proton and neutron diffusenesses to differ.  The Khan prescription includes procedures for determining the neutron and proton radii ($R_n$ and $R_p$) and the neutron and proton diffusenesses ($a_n$ and $a_p$).  Equation (13) from Ref. \cite{Kh22} gives

\begin{equation}
\label{eq:mnmp3}
\frac{M_n}{M_p} = \frac{b_p}{b_n} \frac{a_n}{a_p} \frac{R_p}{R_n} \left[\frac{\delta_{(p,p')}}{\delta_p} \frac{a_p}{a} \left( 1 + \frac{N}{Z} \frac{b_n}{b_p} \right) - 1 \right],
\end{equation}

\noindent
which can be seen to reduce to equation (\ref{eq:mnmp2}) if $a=a_n=a_p$ and $R_p=R_n$. (The ``total" diffuseness $a$ is set to 0.7 fm.) The difference between $R_p$ and $R_n$ depends on $N-Z$, while $a_n$ and $a_p$ depend on the neutron and proton separation energies, respectively.  Khan provides a procedure for determining separation energies, but in the present work we only address stable isotopes so we simply use the empirical separation energies.

Using the Bernstein prescription, Riley \textit{et al.} \cite{Ri25} calculated that in $^{42}$Si $(M_n/M_p)/(N/Z)=0.67(16)$, which was a significant departure from the value of 1.0 expected for a liquid drop.  However, their $(M_n/M_p)/(N/Z)$ result was reproduced by a shell model calculation presented in Ref. \cite{Ri25}.  Riley \textit{et al.} also calculated $(M_n/M_p)/(N/Z)$ using the Khan prescription and obtained the value $(M_n/M_p)/(N/Z)=0.58(28)$, which is once again a significant departure from the value expected in a liquid drop picture.



The 48 isotopes selected for inclusion in the present analysis are listed in Table \ref{tab:long} along with the $(M_n/M_p)/(N/Z)$ results using both the Bernstein and Khan prescriptions.  The electromagnetic matrix elements $B(E2;0_{gs}^+ \rightarrow 2_1^+)$ used in calculating the values in Table \ref{tab:long} are taken from Ref. \cite{Pri16}.  The ($p,p'$) deformation lengths are either those reported in the references listed or are calculated using the reported deformation parameter $\beta_2$, the radius parameter for the real volume potential, $r_0$, and the mass of the nucleus, $A$ using

\begin{equation}
\label{eq:deltapp}
\delta_{(p,p')} = r_0 \beta_2 A^{1/3}.
\end{equation}

For 34 of the 48 nuclei tabulated here, the proton scattering results used to calculate $M_n/M_p$ here were extracted from the differential cross section data using coupled-channels calculations.  These nuclei include the deformed rare earth nuclei for which channel coupling is particularly important.  The nuclei for which only single-step DWBA analyses were performed include $^{40}$Ar, $^{46,48}$Ti, $^{78,80,82}$Kr, $^{138}$Ba, $^{140}$Ce, $^{144}$Sm, $^{194,196,198}$Pt and $^{206,208}$Pb.

To understand the experimental uncertainties in the $\delta_{(p,p')}$ values, we examined several results for $^{40}$Ar.  There are four studies, including the one selected in the present work (Ref. \cite{Bl88}), at beam energies of 20 MeV or higher and which provide results on deformation parameters or deformation lengths in a form that could have been adopted here.  We chose to examine only $^{40}$Ar($p,p'$) studies at beam energies above 20 MeV to eliminate the possibility of compound nucleus effects.  Aside from Ref. \cite{Bl88}, the other three studies of $^{40}$Ar($p,p'$) are the study of De Leo \textit{et al.} at beam energies of 29.6 and 35.1 MeV (Ref. \cite{De85}, published in 1985), the study of Rush \textit{et al.} at beam energies of 30.4 and 49.4 MeV (Ref. \cite{Ru71}, published in 1971) and the study of Johnson and Griffiths at a beam energy of 24.85 MeV (Ref. \cite{Jo68}, published in 1968).  The study by Blanpied \textit{et al.} compiled in Table \ref{tab:long} reports on an experiment with a beam energy of 800 MeV (Ref. \cite{Bl88}, published in 1988).  Blanpied \textit{et al.} used the DWBA (without coupled channels) to obtain a deformation parameter for the $2_1^+$ state in $^{40}$Ar of 0.95 fm.  De Leo \textit{et al.} reported two results from their coupled channels analyses, one assuming a vibrational form factor and the other a rotational form factor.  The analysis with a vibrational form factor yielded $\beta_2=0.242(5)$ (corresponding to $\delta_{(p,p')}=0.943(19)~fm$) while the rotational analysis gave $\beta_2=0.220(4)$ (yielding $\delta_{(p,p')}=0.858(16)~fm$).  With a DWBA analysis, Rush et al. \cite{Ru71} obtained $\beta_2=0.26(2)$ at 30.4 MeV and $\beta_2=0.24(2)$ at 49.4 MeV, corresponding to deformation lengths of $1.014(78)~fm$ and $0.903(75)~fm$, respectively.  Finally, Johnson and Griffiths \cite{Jo68} used a DWBA analysis to obtain $\beta_2=0.21$, which corresponds to $\delta_{(p,p')}=0.847(42)~fm$.  Of these six values, five fall within $7\%$ of the uncertainty-weighted average value of $0.894(11)~fm$.  The result of Rush \textit{et al.} differs from the average by $13\%$. Our conclusion from this study is that we should assign a minimum experimental uncertainly of $7\%$ to deformation lengths from proton scattering studies to account for systematic uncertainties inherent in the model dependence of ($p,p'$) analyses.  That is, if the authors of a study assign an experimental uncertainty smaller than $7\%,$ or if an experimental uncertainty is not reported, we then use $7\%$.  If the authors adopt an experimental uncertainty larger than $7\%$, then we adopt the uncertainty given by the authors.   

\begin{figure}
  \includegraphics[scale=0.62]{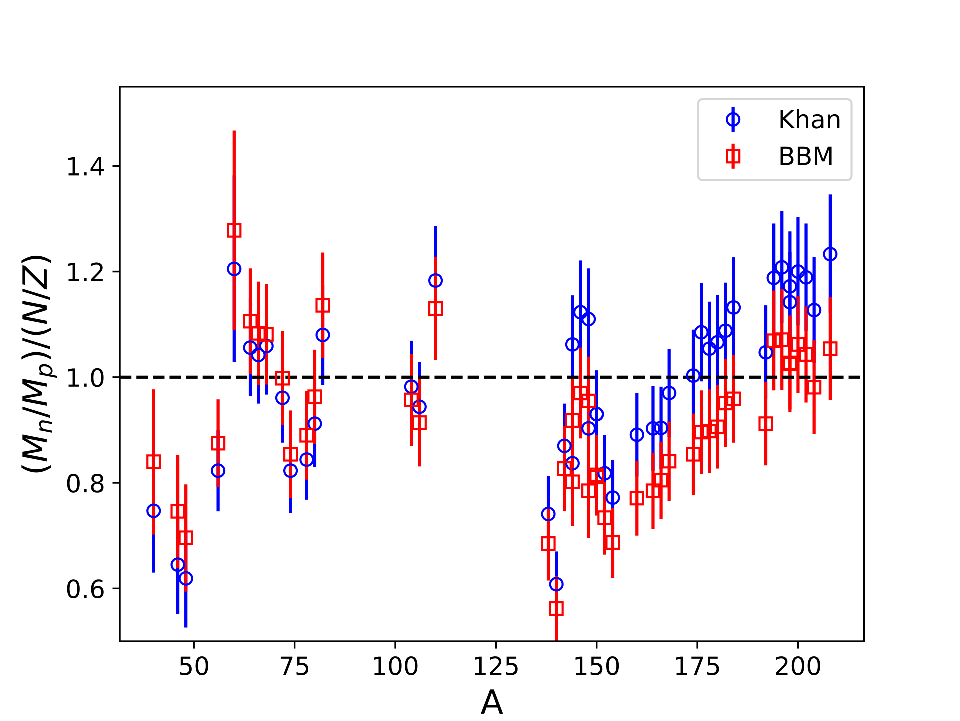}
  \caption{\label{fig:MnMpNZ_A} (Color online) $(M_n/M_p)/(N/Z)$ values calculated with the prescriptions of Refs. \cite{Be83,Kh22} for $0_{gs}^+ \rightarrow 2_1^+$ excitations in the nuclei examined here.}
\end{figure}

The ratios $(M_n/M_p)/(N/Z)$ are displayed in Figure \ref{fig:MnMpNZ_A} for 47 of the 48 isotopes in the table.  The one isotope left out of the figure is $^{206}$Pb, which has extraordinarily large $(M_n/M_p)/(N/Z)$ ratios with both prescriptions.  Including $^{206}$Pb in figure \ref{fig:MnMpNZ_A} would make it more difficult to discern trends for the other nuclei in the figure.   

For all but one of the isotopes examined here, the inelastic proton scattering data used beam energies of 65 MeV or lower.  For these isotopes, we adopted $b_n/b_p=3.00 \pm 0.75$.  The inelastic proton scattering data for the remaining isotope, $^{40}$Ar, was measured at a beam energy of 800 MeV.  For this, we adopted the $b_n/b_p$ value of 0.83 for this energy \cite{Be83}.  Figure \ref{fig:MnMpNZ_A} demonstrates that uncertainties of $7\%$ or greater on proton scattering deformation lengths and the $25\%$ uncertainty on $b_n/b_p$ for low energy proton scattering still allow us to draw useful global conclusions about the results.

The uncertainties in the $(M_n/M_p)/(N/Z)$ values in Table \ref{tab:long} were determined by calculating the impact of each uncertainty (in $\delta_p$, $\delta_{(p,p')}$ and $b_n/b_p$) separately, using upper and lower bounds of the uncertainty range of each parameter in Equations \ref{eq:mnmp2} and \ref{eq:mnmp3}, and combining the resulting contributions in quadrature to find the overall uncertainty.  It is important to note that because $b_n/b_p$ is in both the denominator and part of the numerator in each of Equations \ref{eq:mnmp2} and \ref{eq:mnmp3}, a $25\%$ uncertainty in $b_n/b_p$ does not always yield a correspondingly large uncertainty in the resulting value of $(M_n/M_p)/(N/Z)$.

\begin{center}
    
\begin{longtable}{|l|l|l|l|l|}
\caption{\label{tab:mnmp} Nuclei included in the $M_n/M_p$ analysis. The electromagnetic data are taken from Ref. \cite{Pri16}.}  \label{tab:long} \\

\hline \multicolumn{1}{|c|}{Nucleus} & \multicolumn{1}{c|}{$\delta_{(p,p')}$ (fm)} & \multicolumn{1}{c|}{Ref. for ($p,p'$) data} & \multicolumn{1}{c|}{$(M_n/M_p)/(N/Z)$ Ref. \cite{Be83}} & \multicolumn{1}{c|}{$(M_n/M_p)/(N/Z)$ Ref. \cite{Kh22}} \\ \hline 
\endfirsthead

\multicolumn{5}{c}%
{{\tablename\ \thetable{} -- continued from previous page}} \\
\hline \multicolumn{1}{|c|}{Nucleus} & \multicolumn{1}{c|}{$\delta_{(p,p')}$ (fm)} & \multicolumn{1}{c|}{Ref. for ($p,p'$) data} & \multicolumn{1}{c|}{$(M_n/M_p)/(N/Z)$ Ref. \cite{Be83}} & \multicolumn{1}{c|}{$(M_n/M_p)/(N/Z)$ Ref. \cite{Kh22}}\\ \hline 
\endhead

\hline \multicolumn{5}{|c|}{{Continued on next page}} \\ \hline
\endfoot

\hline \hline
\endlastfoot

$^{40}$Ar & 0.95(7) & \cite{Bl88} & $0.84(14)$ & $0.75(12)$ \\
$^{46}$Ti & 1.10(11) & \cite{Fu87} & $0.75(11)$ & $0.65(9)$ \\
$^{48}$Ti & 0.86(9) & \cite{Hi89} & $0.70(10)$ & $0.62(9)$ \\
$^{56}$Fe & 0.99(7) & \cite{deL96} & $0.88(8)$ & $0.82(8)$ \\
$^{60}$Ni & 1.17(14) & \cite{va89} & $1.28(19)$ & $1.21(18)$ \\
$^{64}$Zn & 1.22(8) & \cite{Ja87a} & $1.11(10)$ & $1.06(9)$ \\
$^{66}$Zn & 1.13(8) & \cite{Ja87a} & $1.08(10)$ & $1.04(9)$ \\
$^{68}$Zn & 1.05(7) & \cite{Ja87b} & $1.08(9)$ & $1.06(9)$ \\
$^{72}$Ge & 1.20(8) & \cite{Mo93} & $1.00(9)$ & $0.96(8)$ \\
$^{74}$Ge & 1.27(9) & \cite{Mo93} & $0.85(8)$ & $0.82(8)$ \\
$^{78}$Kr & 1.65(12) & \cite{Sa79} & $0.89(8)$ & $0.84(8)$ \\
$^{80}$Kr & 1.35(9) & \cite{Sa79} & $0.96(9)$ & $0.91(8)$ \\
$^{82}$Kr & 1.17(8) & \cite{Sa79} & $1.14(10)$ & $1.08(9)$ \\
$^{104}$Pd & 1.13(8) & \cite{Pi90} & $0.96(9)$ & $0.98(9)$ \\
$^{106}$Pd & 1.21(8) & \cite{Pi90} & $0.91(8)$ & $0.94(8)$ \\
$^{110}$Pd & 1.53(11) & \cite{Pi90} & $1.13(10)$ & $1.18(10)$ \\
$^{138}$Ba & 0.43(3) & \cite{La74} & $0.69(7)$ & $0.74(7)$ \\
$^{140}$Ce & 0.41(3) & \cite{Sh77} & $0.56(6)$ & $0.61(6)$ \\
$^{142}$Nd & 0.49(4) & \cite{Pi93} & $0.83(8)$ & $0.87(8)$ \\
$^{144}$Nd & 0.74(5) & \cite{Pi93} & $0.92(8)$ & $1.06(9)$ \\
$^{146}$Nd & 0.93(6) & \cite{Pi93} & $0.97(9)$ & $1.12(10)$ \\
$^{148}$Nd & 1.22(9) & \cite{Pi93} & $0.96(8)$ & $1.11(10)$ \\
$^{150}$Nd & 1.52(11) & \cite{Pi93} & $0.81(7)$ & $0.93(8)$ \\
$^{144}$Sm & 0.46(3) & \cite{La74} & $0.80(8)$ & $0.84(8)$ \\
$^{148}$Sm & 0.74(6) & \cite{Co93} & $0.79(9)$ & $0.90(10)$ \\
$^{150}$Sm & 1.05(7) & \cite{Pi90} & $0.82(7)$ & $0.93(8)$ \\
$^{152}$Sm & 1.54(11) & \cite{Ic87} & $0.73(7)$ & $0.82(7)$ \\
$^{154}$Sm & 1.63(11) & \cite{Ic87} & $0.69(7)$ & $0.77(7)$ \\
$^{160}$Gd & 1.86(13) & \cite{Ic87} & $0.77(7)$ & $0.89(8)$ \\
$^{164}$Dy & 1.89(13) & \cite{Ic87} & $0.79(7)$ & $0.90(8)$ \\
$^{166}$Er & 1.88(13) & \cite{Ic87} & $0.81(7)$ & $0.90(8)$ \\
$^{168}$Er & 1.94(14) & \cite{Ic87} & $0.84(7)$ & $0.97(8)$\\
$^{174}$Yb & 1.90(13) & \cite{Ic84} & $0.85(8)$ & $1.00(9)$ \\
$^{176}$Yb & 1.85(13) & \cite{Ic87} & $0.90(8) $ & $1.09(9)$ \\
$^{178}$Hf & 1.72(12) & \cite{Og86} & $0.90(8)$ & $1.05(9)$ \\
$^{180}$Hf & 1.71(12) & \cite{Og86} & $0.91(8)$ & $1.07(9)$\\
$^{182}$W & 1.62(11) & \cite{Og86} & $0.95(8)$ & $1.09(9)$ \\
$^{184}$W & 1.54(11) & \cite{Og86} & $0.96(8)$ & $1.13(9)$ \\
$^{192}$Os & 1.05(7) & \cite{Ic87} & $0.91(8)$ & $1.05(9)$ \\
$^{194}$Pt & 1.04(7) & \cite{De81} & $1.07(9)$ & $1.19(10)$ \\
$^{196}$Pt & 0.96(7) & \cite{De81} & $1.07(9)$ & $1.21(11)$ \\
$^{198}$Pt & 0.81(6) & \cite{De81} & $1.03(9)$ & $1.17(10)$ \\
$^{198}$Hg & 0.75(5) & \cite{Ho91} & $1.03(9)$ & $1.14(9)$ \\
$^{200}$Hg & 0.72(5) & \cite{Ho91} & $1.06(9)$ & $1.20(10)$ \\
$^{202}$Hg & 0.60(4) & \cite{Ho91} & $1.04(9)$ & $1.19(10)$ \\
$^{204}$Hg & 0.47(3) & \cite{Ho91} & $0.98(9)$ & $1.13(10)$ \\
$^{206}$Pb & 0.46(3) & \cite{Fi83} & $2.24(19)$ & $2.48(20)$ \\
$^{208}$Pb & 0.40(3) & \cite{Wa75} & $1.05(10)$ & $1.23(11)$ \\

\end{longtable}
\end{center}

There are several interesting aspects to Fig. \ref{fig:MnMpNZ_A}.  First, it demonstrates that the two prescriptions for calculating $M_n/M_p$ \cite{Be83,Kh22} give the same general behavior up to the Pd isotopes.  However, at masses above 140, the differences between $R_n$ and $R_p$, which are driven by $N-Z$, result in significant differences between the results of the two prescriptions.

Second, there are three mass regions in which $(M_n/M_p)/(N/Z)$ is systematically below 1.0.  One of those regions includes $^{40}$Ar, $^{46,48}$Ti and $^{56}$Fe, and the second includes the $N=82$ isotones $^{138}$Ba, $^{140}$Ce, $^{142}$Nd and $^{144}$Sm.  The third such region includes the deformed rare earth nuclei $^{150}$Nd, $^{150,152,154}$Sm, $^{160}$Gd, $^{164}$Dy and $^{166}$Er.  

We begin by examining the $N=82-126$ nuclei, which includes both the $N=82$ semi-magic isotones and the deformed rare earth nuclei listed in the previous paragraph.  Figure \ref{fig:MnMpNZ_A} shows that the $N=82$ isotones have $(M_n/M_p)/(N/Z)$ values significantly smaller than 1.0.  This is exactly what would be expected for closed-neutron-shell nuclei - that the protons play a disproportionately large role in the excitations because the neutron contributions are suppressed by the neutron shell closure.  Then the $N=84$, $86$ and $88$ vibrational and transitional nuclei $^{144,146,148}$Nd have $(M_n/M_p)/(N/Z)$ values that are consistent with or close to 1.0, as would be expected for such nuclei.  However, the $N=86$ transitional nucleus $^{148}$Sm has an $(M_n/M_p)/(N/Z)$ value which is significantly below 1.0.

One notable feature in the $N=82-126$ region is that the $(M_n/M_p)/(N/Z)$ values in the well-deformed $N \ge 90$ Nd, Sm, Gd, Dy and Er isotopes are systematically significantly lower than 1.0.  That is, the proton transition matrix elements $M_p$ in these deformed rare earth nuclei are disproportionately large - a behavior that cannot be explained in the context of the simple liquid drop model that assumes that protons and neutrons are homogeneously distributed in the nucleus.

\begin{figure}
 \includegraphics[scale=0.62]{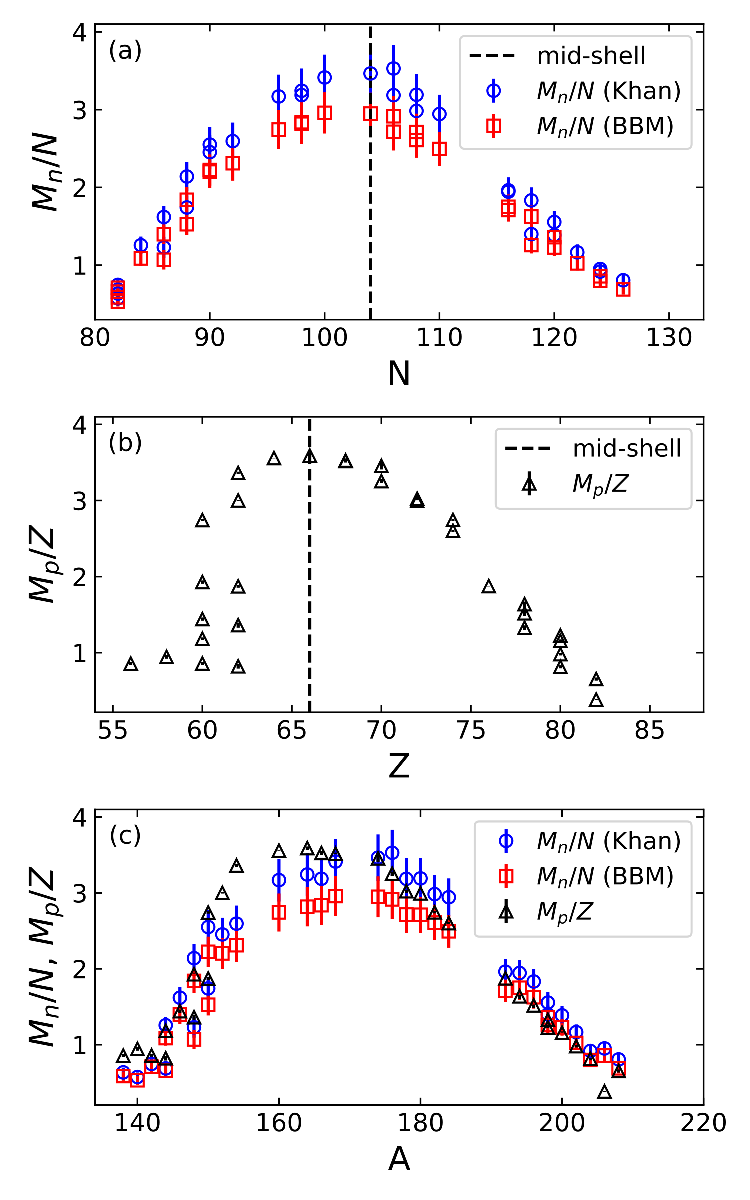}
  \caption{\label{fig:N82_126} (Color online) Neutron and proton transition matrix elements per nucleon for the $0_{gs}^+ \rightarrow 2_1^+$ transitions in the $N=82-126$ nuclei listed in Table \ref{tab:long}.  Panel (a) shows only the neutron transition matrix elements (but calculated with both prescriptions from Refs. \cite{Be83,Kh22}), with the midshell at $N=104$ noted.  Panel (b) shows only proton transition matrix elements with the midshell at $Z=66$ noted.  Panel (c) shows both neutron and proton transition matrix elements plotted against $A$.}    
\end{figure}

However, these results in the $N=82-126$ can be explained using a schematic picture in which the neutron and proton fluids are each most deformed at the midshell of their species, which is $Z=66$ for protons and $N=104$ for neutrons.  The first two panels of Figure \ref{fig:N82_126} illustrate the transition matrix element per nucleon for each species.  The values of $M_n$ shown in panel (a) are simply calculated by multiplying the $M_n/M_p$ values by $M_p$ values deduced from the $B(E2;0_{gs}^+ \rightarrow 2_1^+)$ values compiled in Ref. \cite{Pri16} using 

\begin{equation}
\label{eq:mp}
M_p^2 = B(E2;0_{gs}^+ \rightarrow 2_1^+).
\end{equation}

The quantity $M_n/N$ reaches a maximum at or near the midshell point $N=104$, while $M_p/Z$ reaches a maximum at or near the midshell point of $Z=66$.  The valley of stability reaches the proton midshell ($Z=66$) before it reaches the neutron midshell ($N=104$).  Therefore, $M_n/M_p$ is significantly smaller than the liquid drop expectation of $N/Z$ in several nuclei around $^{164}$Dy.  Once the neutron number reaches 104, which occurs along the valley of stability at $^{174}$Yb, $M_n/M_p$ values have recovered to be near or equal to $N/Z$.       

We would have expected that as nuclei that are apparently vibrational, $^{40}$Ar, $^{46,48}$Ti and $^{56}$Fe would have had $(M_n/M_p)/(N/Z)$ values near 1.0.  However, each of these four nuclei have values significantly smaller than 1.0.  It is important to note that recent IMSRG calculations \cite{Ya18,Ya20} suggest that $^{48}$Ti has a stable quadrupole deformation.  The $B(E2;0_{gs}^+ \rightarrow 2_1^+)$ strength in $^{46}$Ti is very close to that in $^{48}$Ti \cite{Pri16,Gr24}, so $^{46}$Ti might have a stable quadrupole deformation as well.  In any event, it would be useful to see if a shell model calculation could reproduce the $0_{gs}^+ \rightarrow 2_1^+$ strengths of all four of these isotopes.    

The $(M_n/M_p)/(N/Z)$ results for $^{206}$Pb for both the Ref. \cite{Be83} and Ref. \cite{Kh22} prescriptions provide a textbook example of the behavior of the $0_{gs}^+ \rightarrow 2_1^+$ excitation in an SCS nucleus.  With a closed proton shell, $^{206}$Pb should have $(M_n/M_p)/(N/Z)>N/Z$, and the two empirical values, $2.24(15)$ and $2.48(17)$, are the largest we see in our compilation by far.  
The other closed-proton-shell nucleus in our compilation, $^{60}$Ni, also has $(M_n/M_p)/(N/Z)>1.0$ for both prescriptions [1.28(19) and 1.21(18) using the prescriptions of Refs. \cite{Be83} and \cite{Kh22}, respectively]. However, the uncertainties are large.

One way to expand the present study would be to compare inelastic $\alpha$-particle scattering data, for which $b_n/b_p=1$, to electromagnetic data to see whether the conclusions reached with such comparisons are consistent with those reached in the present work.  Another way to extend the present study would be to compare inelastic deuteron scattering data, which once again has $b_n/b_p=1$, to electromagnetic data.  There are several deformed actinide nuclei for which inelastic deuteron scattering data are available, and looking for deviations from $M_n/M_p=N/Z$ in those nuclei would provide an additional perspective on the present results. 

In summary, we have examined $(M_n/M_p)/(N/Z)$ values for $0_{gs}^+ \rightarrow 2_1^+$ transitions in 48 stable even-even nuclei in which electromagnetic matrix elements are available in the compilation of Ref. \cite{Pri16} and for which the ratios $(M_n/M_p)/(N/Z)$ can be extracted from high-quality inelastic proton scattering data.  The deviation of a number of deformed rare earth nuclei from the value of 1.0 that would be expected from the simple liquid drop model can be explained in a schematic picture in which $M_p$ reaches its maximum value at proton midshell ($Z=66$) and $M_n$ reaches its maximum value at neutron midshell ($N=104$).  Shell model calculations with modern interactions would likely provide insight about why the vibrational mid-mass nuclei $^{40}$Ar, $^{46,48}$Ti and $^{56}$Fe all have $(M_n/M_p)/(N/Z)$ values significantly smaller than 1.0.  The results on stable nuclei discussed here, when taken together with the results on the radioactive nuclei discussed in Ref. \cite{Ri25}, demonstrate that the determination of proton and neutron transition matrix elements $M_p$ and $M_n$ using complementary experimental probes can lead to insights that cannot be accessed with electromagnetic measurements of transitions alone.  Therefore, such measurements should be included in present and future investigations of exotic nuclei.    

  
PDC wishes to thank Kun Yang, Jorge Piekarewicz and Alexander Volya for very helpful discussions.  This work was supported by National Science Foundation grants PHY-2412808 and PHY-2208804 and by the US Department of Energy through awards DE-SC0009883 and DE-SC0023633.


\providecommand{\noopsort}[1]{}\providecommand{\singleletter}[1]{#1}%

\end{document}